\documentclass[preprint,12pt]{elsarticle}
\usepackage{amsmath}
\usepackage{amsfonts}
\usepackage{amssymb}
\usepackage{graphicx}
\usepackage{bm}
\numberwithin{equation}{section}
\begin{document}
\begin{frontmatter}
\title{On the coherent states for a relativistic scalar particle}
\author[]{K. Kowalski\corref{cor1}}
%\ead{kowalski@uni.lodz.pl}
%\cortext[cor1]{Corresponding author.}
\author{J. Rembieli\'nski}
\address{Department of Theoretical Physics, University
of \L\'od\'z, ul.\ Pomorska 149/153, 90-236 \L\'od\'z,
Poland}
\author{J.-P. Gazeau}
\address{APC, UMR 7164, Univ Paris  Diderot, Sorbonne Paris Cit\'e  
75205 Paris, France}
\begin{abstract}
The three approaches to relativistic generalization of coherent
states are discussed in the simplest case of a spinless particle: 
the standard, canonical coherent states, the Lorentzian states and the coherent states
introduced by Kaiser and independently by Twareque Ali, Antoine and Gazeau.
All treatments utilize the Newton-Wigner localization and dynamics described
by the Salpeter equation.  The behavior of expectation values of relativistic 
observables in the coherent states is analyzed in detail and the Heisenberg 
uncertainty relations are investigated.
\end{abstract}
\begin{keyword}
Quantum mechanics, relativistic quantum mechanics; coherent states  
\end{keyword}
\end{frontmatter}
\section{Introduction}
In spite of the fact that coherent states are of fundamental importance in quantum 
physics, the problem of their relativistic generalization remains open.  
The difficulties in extending the concept of coherent states to the relativistic 
domain are related with problems in constructing relativistic quantum mechanics.
In particular, they are connected with identification of the appropriate relativistic 
counterpart of the Schr\"odinger equation and corresponding position and 
momentum operators acting on a concrete Hilbert space. The most popular relativistic 
generalizations of the Schr\"odinger equation for the scalar massive particles are
the Klein-Gordon equation \cite{1,2,3,4} and the less known ``square root" of this equation usually
referred to as the Salpeter equation \cite{5,6,7,8,9,10,11,12,13,14,15,16,17,18,19,20,21,22,23,24,
25,26,27,28}.  We remark that the latter equation appears in the context of the so called 
relativistic Hamiltonian dynamics \cite{29,30}.  An asset
of the Klein-Gordon equation is its manifest covariance.  The grave flaw of this equation
is the problem with probablistic interpretation, namely the probability density can be
negative.  To circumvent the problem the probability current is reinterpreted as a
charge current.  Furthermore, we deal with paradoxes such as ``Zitterbewegung" and Klein
paradox, connected with existence of solutions corresponding to negative energy.
To bypass the problem one usually suggests the creation of particle-antiparticle pairs
and conclude that there is no consistent one-particle relativistic quantum mechanics.
The great advantage of the Salpeter equation is that it has well-defined probability density
and current \cite{25}.  Recently, it has been demonstrated \cite{31} that this current is consistent 
with relativistic generalization of the Wigner function \cite{32}.  Furthermore, the Salpeter
equation possesses only positive energy solutions, so the paradoxes do not occur such as
``Zitterbewegung" and Klein paradox mentioned earlier.  Its flaw is that it 
is not manifestly covariant.  Another problem that is sometimes pointed 
out is the nonlocality of the Hamiltonian referring to the Salpeter 
equation which is the pseudodifferential operator.  In spite of its 
nonlocality, the Salpeter equation does not disturb the light cone 
structure and the second quantized version of this equation is macrocausal 
\cite{12}.  Despite its flaws, the advantages of the Salpeter equation were 
the reason for its wide usage in the description of the quark-antiquark 
gluon system \cite{33,34}.

As far as we are aware the first example of relativistic coherent states in the 
literature were the states for a spin 0 and spin-1/2 charged particle in an external 
magnetic field discussed by Malkin and Man'ko in \cite{35}, and by Lev, Semenov, Usenko
and Klauder in more recent papers \cite{36,37,38}.  The relativistic 
dynamics chosen in these papers is described by the Klein-Gordon equation 
or its matrix first-order counterpart introduced by Feshbach and Villars \cite{39}.
The problem with the introduced coherent states is, among others, their 
interpretation as the states closest to the classical ones.  This issue is explicitly
recognized in ref.\ 35.  As a matter of fact the uncertainty relations were discussed
in \cite{38} where the Klein-Gordon equation on a ``null" plane was utilized.  Nevertheless,
the condition for minimalization of these relations in coherent states found in \cite{38}
mixes temporal and spatial coordinates, so it is time-dependent in contrary to the
fact that definition of coherent states should not depend on dynamics --- 
they should be parametrized only by points of the phase space. Yet another 
example of relativistic coherent states based on the Klein-Gordon equation 
are the states introduced in [40].

In this work we discuss the coherent states for a spinless particle assuming the 
dynamics is described by the spinless Salpeter and the position and momentum 
operators have the corresponding $L^2$ representation.  We study the three 
candidates for coherent states in relativistic quantum mechanics.  The first ones are 
the standard, canonical coherent states that were not investigated in the relativistic context so far.
We also study the states related to the exact solutions to the 
Salpeter equation referred in the massless case as to ``Lorentzian'' 
packages and the states introduced by Bakke and Wergeland \cite{41,42} and
Kaiser \cite{43} as well as independently by Twareque Ali, Antoine and Gazeau 
\cite{44,45}.  We obtain, among others, the new formulas related to these coherent 
states and discuss the corresponding uncertainty relations. 
\section{Preliminaries}
In this section we briefly sketch some basic facts about the Salpeter 
equation and the corresponding Hilbert space (see for example \cite{23,25}).
Consider the Hamiltonian of a free relativistic classical particle 
\begin{equation}
%<2.1>
H=\sqrt{c^2{\bm p}^2+m^2c^4},
\end{equation}
where $m$ is the mass of a particle, ${\bm x}$ is its three-position and $c$ 
is the speed of light.  Applying to (2.1) the quantization procedure based 
on the Newton-Wigner localization scheme \cite{6} using the standard 
Schr\"odinger quantization rule $\hat{\bm x}\to{\bm x}$, and 
$\hat{\bm p}\to-{\rm i}\hbar{\bm\nabla}$, we arrive at the Salpeter 
equation such that
\begin{equation}
%<2.2>
{\rm i}\hbar\frac{\partial\phi({\bm x},t)}{\partial t}
=\sqrt{-\hbar^2c^2\Delta+m^2c^4}\phi({\bm x},t),
\end{equation}
where the space of solutions to the Salpeter equation is $L^2(\mathbb{R}^3,d^3{\bm x})$
specified by the scalar product
\begin{equation}
%<2.3>
\langle \phi|\psi\rangle = \int d^3{\bm x}\,\phi^*({\bm x})\psi({\bm x}).
\end{equation}
We point out that every solution of the Salpeter equation (2.2) is also 
the solution of the free Klein-Gordon equation.  The opposite is clearly 
not true because of existence of negative energy solutions to the Klein-Gordon
equation. For a more complete discussion of relations between the Salpeter and 
Klein-Gordon equation we refer the reader to \cite{25}.

On performing the Fourier transformation
\begin{equation}
%<2.4>
\phi({\bm x},t)=\frac{1}{(2\pi)^\frac{3}{2}\hbar^3}\int d^3{\bm
p}\,e^{{\rm i}\frac{{\bm p}\mbox{\boldmath$\scriptstyle{\cdot}$}
{\bm x}}{\hbar}}\tilde\phi({\bm p},t),
\end{equation}
we get the Salpeter equation in the momentum representation
\begin{equation}
%<2.5>
{\rm i}\hbar\frac{\partial\tilde\phi({\bm p},t)}{\partial t}=\sqrt
{{\bm p}^2c^2+m^2c^4}\phi({\bm p},t),
\end{equation}
where the space of solutions to (2.5) is the Hilbert space 
$L^2(\mathbb{R}^3,d^3{\bm p})$ with the scalar product
\begin{equation}
%<2.6>
\langle\phi|\psi\rangle = \frac{1}{\hbar^3}\int d^3{\bm p}\,{\tilde\phi}^*({\bm p})
{\tilde\psi}({\bm p}).
\end{equation}
Clearly, the action of position and momentum operators is given by the 
standard Schr\"odinger rule
\begin{equation}
%<2.7>
\hat{\bm x}\tilde\phi({\bm p})={\rm i}\hbar{\bm\nabla}_{{\bm 
p}}\tilde\phi({\bm p}),\qquad\hat{\bm p}\tilde\phi({\bm p})=
{\bm p}\tilde\phi({\bm p}).
\end{equation}
We remark that the measure $d^3{\bm p}$ is not invariant measure on a 
mass hyperboloid.  The isomorphism of the Hilbert space specified by the 
scalar product (2.6) and the Hilbert space with the invariant measure 
$\frac{mc\,d^3{\bm p}}{p^0({\bm p})}$ such that
\begin{equation}
%<2.8>
\langle\phi'|\psi'\rangle = \frac{1}{\hbar^3}\int\frac{mc\,d^3{\bm p}}{p^0({\bm p})}
{\tilde\phi}'^*({\bm p}){\tilde\psi}'({\bm p}),
\end{equation}
where $p^0({\bm p})=\sqrt{{\bm p}^2+m^2c^2}$, is given by the 
transformation
\begin{equation}
%<2.9>
\tilde\phi'({\bm p})=\frac{1}{mc}\sqrt{p^0({\bm p})}\tilde\phi({\bm p}).
\end{equation}
The position and momentum operators act in the Hilbert space with the 
scalar product (2.8) in the following way
\begin{equation}
%<2.10>
\hat{\bm x}_{\rm NW}\tilde\phi'({\bm p})=\left({\rm i}\hbar{\bm\nabla}_{{\bm 
p}}-\frac{{\rm i}\hbar}{2}\frac{{\bm p}}{{\bm p}^2+m^2c^2}\right)\tilde
\phi'({\bm p}),\qquad\hat{\bm p}\tilde\phi'({\bm p})={\bm p}\tilde\phi'({\bm p}).
\end{equation}
We point out that our choice of rather nonstandard form of the invariant 
measure was motivated by limiting procedures in the nonrelativistic case and
some dimensional deliberations.  The position operator from the formula (2.10) 
is the well known Newton-Wigner operator. It should be noted that the Newton-Wigner 
operator is not covariant.
\section{Standard coherent states in relativistic quantum mechanics}
\subsection{Standard coherent states}
An experience with construction of the standard, canonical coherent states based on
canonical commutation relations $[a,a^\dagger]=I$, and their 
numerous generalizations such as for example coherent states for a particle 
on a circle, sphere and torus \cite{46,47,48,49} shows that one of their 
most important property is not a temporal stability with respect to some 
particular evolution but the parametrization by points of the classical phase 
space.  On the other hand, we have the standard $L^2$ Schr\"odinger realization 
of the space of solutions to the Salpeter equation described in the previous 
section.  Therefore, it seems plausible to discuss the possibility of applying 
the canonical coherent states also in relativistic case.  Of course the canonical
coherent states are not Lorentz covariant and should be defined in a concrete reference
frame.  Nevertheless, it is interesting to compare the behavior of average of relativistic
observables in these states and covariant ones.  Consider the canonical 
coherent states.  We restrict for simplicity to the one-dimensional configuration 
space.  These states are eigenvectors of the annihilation operator
\begin{equation}
%<3.1>
\hat a|z\rangle=z|z\rangle,
\end{equation}
where
\begin{equation}
%<3.2>
\hat a=\frac{1}{\sqrt{2}}\left(\frac{\hat x}{\sigma}+{\rm i}\frac{\sigma\hat p}
{\hbar}\right)
\end{equation}
and $\sigma$ is a constant with dimension of length. In the case of the 
harmonic oscillator coherent states we have $\sigma=\sqrt{\hbar/m\omega}$. 
The complex number $z$ labelling the coherent states can be expressed by 
means of the average position $\bar x=\langle z|\hat x|z\rangle$ and 
momentum $\bar p=\langle z|\hat p|z\rangle$ in the form
\begin{equation}
%<3.3>
z=\frac{1}{\sqrt{2}}\left(\frac{\bar x}{\sigma}+{\rm i}\frac{\sigma\bar p}
{\hbar}\right).
\end{equation}
The standard deviations in the coherent state $|z\rangle$ satisfy
\begin{equation}
%<3.4>
\Delta\hat x =\frac{\sigma}{\sqrt{2}},\qquad \Delta 
\hat p=\frac{\hbar}{\sqrt{2}\sigma},\qquad \Delta\hat x \Delta\hat 
p=\frac{\hbar}{2},
\end{equation}
so the coherent states minimalize the Heisenberg uncertainty relations.  
In this sense the coherent states are closest to the classical ones.  The 
coherent states are not orthogonal.   We have
\begin{equation}
%<3.5>
\langle z|w\rangle=\exp[-\textstyle{\frac{1}{2}}(|z|^2+|w|^2-2z^*w)].
\end{equation}
The coherent states form the complete (overcomplete) set.  Namely,
\begin{equation}
%<3.6>
\frac{1}{\pi}\int d{\rm Re}z\,d{\rm Im}z\,|z\rangle\langle z|=I,
\end{equation}
or equivalently
\begin{equation}
%<3.7>
\frac{1}{2\pi\hbar}\int d\bar x d\bar p\,|\bar x,\bar p\rangle
\langle\bar x,\bar p|=I, 
\end{equation}
where $|\bar x,\bar p\rangle\equiv|z\rangle$, with $z$ expressed by (3.3).

Consider now the one-dimensional counterpart of the momentum 
representation specified by (2.6) and (2.7)
\begin{align}
\langle\phi|\psi\rangle &=\frac{1}{\hbar}\int 
dp\,{\tilde\phi}^*(p)\tilde\psi(p),\\
\hat x\tilde\phi(p) &={\rm i}\hbar\frac{\partial\tilde\phi}{\partial p},\qquad 
\hat p\tilde\phi(p)=p\tilde\phi(p),
\end{align}
where $\tilde\phi(p)=\langle p|\phi\rangle$, and the vectors $|p\rangle$ span the 
momentum representation.
 
The normalized coherent states in the momentum representation are given by
\begin{align}
\tilde\phi_z(p) &=\frac{\sigma^{\frac{1}{2}}}{\pi^{\frac{1}{4}}}
e^{\textstyle{\frac{1}{2}}(z^2-|z|^2)}e^{-\textstyle{\frac{1}{2}}
\frac{\sigma^2p^2}{\hbar^2}-\frac{{\rm i}\sqrt{2}z\sigma p}{\hbar}},\\
\tilde\phi_z(p) &=\frac{\sigma^{\frac{1}{2}}}{\pi^{\frac{1}{4}}}
\exp\left[-\frac{\sigma^2}{2\hbar^2}(p-\bar p)^2-{\rm i}\frac{{\bar x} 
p}{\hbar}\right],
\end{align}
where $\tilde\phi_z(p)=\langle p|z\rangle\equiv\langle p|\bar x,\bar 
p\rangle$, and $z$ is given by (3.3).

We finally write down the formula for the average value of the energy of the
free nonrelativistic particle such that
\begin{equation}
%<3.12>
\bar E=\langle z|\frac{{\hat p}^2}{2m}|z\rangle=\frac{{\bar 
p}^2}{2m}+\frac{\hbar^2}{4\sigma^2m}.
\end{equation}
Thus, up to additive constant, we have the classical relation between the average 
energy and momentum of a particle.  An experience with coherent states for 
a particle on a circle and sphere \cite{46,47} shows that such 
behavior is one of characteristic properties of coherent states.
\subsection{Averages of relativistic observables in the canonical coherent states}
We now investigate the closeness of the canonical coherent states to the 
classical phase space in the relativistic case by means of the analysis of 
averages of relativistic observables in these states.  Consider the 
Hamiltonian of a free relativistic particle
\begin{equation}
%<3.13>
H=\sqrt{c^2{\hat p}^2+m^2c^4}.
\end{equation}
Using (3.8), (3.9) and (3.11) we find that the average energy in the 
canonical coherent state is
\begin{equation}
%<3.14>
\bar E=\langle z|\sqrt{c^2{\hat p}^2+m^2c^4}|z\rangle=\frac{\sigma}{\sqrt{\pi}\hbar}
\int_{-\infty}^\infty e^{-\frac{\sigma^2}{\hbar^2}(p-\bar p)^2}\sqrt{c^2 p^2+m^2c^4}\,dp.
\end{equation}
Expanding the exponential function in (3.14) into the power series and using
the identity
\begin{equation}
%<3.15>
\int_{-\infty}^\infty e^{-\beta x^2}\sqrt{x^2+\gamma^2}\,dx = \frac{\sqrt{\pi}}{\beta}
U(-\textstyle{\frac{1}{2}},0,\beta\gamma^2),\quad {\rm Re}\beta>0,\quad 
{\rm Re}\gamma^2>0,
\end{equation}
where $U(a,b,z)$ is the confluent hypergeometric function \cite{50}, 
and (A2) and (A3), we get the following power series representation of the 
integral from (3.14) 
\begin{equation}
%<3.16>
\bar E=\frac{c\hbar}{\sigma}e^{-\left(\frac{\sigma\bar p}{\hbar}\right)^2}
\sum_{n=0}^{\infty}\frac{1}{n!}\left(\frac{\sigma\bar 
p}{\hbar}\right)^{2n}U\left[-\frac{1}{2},-n,\left(\frac{\sigma}{\lambda_c}\right)^2\right],
\end{equation}
where $\lambda_c=\frac{\hbar}{mc}$ is the Compton wavelength.
The numerical calculations based on (3.14) or (3.16) show that for 
$\sigma/\lambda_c\geqslant5$ (the larger fraction the better approximation) 
we have an approximate relation
\begin{equation}
%<3.17>
\bar E\approx \sqrt{{\bar p}^2c^2+m^2c^4},
\end{equation}
where the approximation is very good.  Namely, for $\sigma/\lambda_c=5$ the maximal 
relative error $(\bar E-\sqrt{{\bar p}^2c^2+m^2c^4})/\bar E$ arising 
for $\bar p=0$ is of order 1\%.  We point out that $\sigma=5\lambda_c$ 
is still the length scale when the description is adequate based on relativistic 
quantum mechanics.  This means that the parameter $\bar p$ labelling the coherent states 
can be regarded as the classical momentum also in the relativistic case. 

We now discuss the case of a massless relativistic particle.  The 
Hamiltonian (3.13) takes the form
\begin{equation}
%<3.18>
H=c|\hat p|.
\end{equation}
Taking into account (3.8) and (3.11) we get
\begin{equation}
%<3.19>
\bar E=\langle z|c|\hat p||z\rangle=\frac{\sigma}{\sqrt{\pi}\hbar}
\int_{-\infty}^\infty |p|e^{-\frac{\sigma^2}{\hbar^2}(p-\bar p)^2}dp.
\end{equation}
Hence, using the identity \cite{51}
\begin{align}
%<3.20>
\int_0^{\infty}xe^{-\mu x^2-2\nu x}dx =&\frac{1}{2\mu}-\frac{\nu}{2\mu}
\sqrt{\frac{\pi}{\mu}}e^{\frac{\nu^2}{\mu}}\left[1-{\rm erf}
\left(\frac{\nu}{\sqrt{\mu}}\right)\right],\nonumber\\
&\Big\vert{\rm arg}\nu<\frac{\pi}{2}\Big\vert,\quad {\rm Re}\mu>0,
\end{align}
where ${\rm erf}(x)$ is the error function, we find
\begin{equation}
%<3.21>
\bar E=c{\bar p}{\rm erf}\left(\frac{\sigma\bar p}{\hbar}\right)+
\frac{c\hbar}{\sqrt{\pi}\sigma}e^{-\left(\frac{\sigma\bar p}{\hbar}\right)^2}.
\end{equation}
Since ${\rm erf}(x)$ is an odd function we have
\begin{equation}
%<3.22>
\bar E=c|{\bar p}|{\rm erf}\left(\frac{\sigma|\bar p|}{\hbar}\right)+
\frac{c\hbar}{\sqrt{\pi}\sigma}e^{-\left(\frac{\sigma\bar p}{\hbar}\right)^2}.
\end{equation}
From numerical simmulations we find that whenever $\sigma/{\bar\lambda}\geqslant0.21$, 
where $|\bar p|=h/{\bar\lambda}$, and $h=2\pi\hbar$ is the Planck constant, which leads 
to $\bar\lambda\leqslant4.76\sigma$, then we have
\begin{equation}
%<3.23>
\bar E\approx c|\bar p|.
\end{equation}
where the approximation is very good. Therefore, the parameter $\bar p$ marking the 
coherent states can be interpreted as the classical momentum of a massless particle.  
We point out that the condition $\bar\lambda\leqslant4.76\sigma$ implies for the 
value of $\sigma$ analogous to that applied in the massive case such that
$\sigma=5\lambda_c^{(\pi^+)}$, where $\lambda_c^{(\pi^+)}$ is the Compton 
wavelength of pion $\pi^+$, the estimate $\bar\lambda\leqslant 
4{\cdot}10^{-5}\,{\rm nm}$.  The value $4{\cdot}10^{-5}\,{\rm nm}$ 
corresponds to gamma rays.  This result seems to be reasonable from the 
physical point of view.

Consider finally the expectation value of the operator of the relativistic 
velocity
\begin{equation}
%<3.24>
\hat v=\frac{c\hat p}{\sqrt{{\hat p}^2+m^2c^2}}.
\end{equation}
On taking into account (3.8) and (3.11) we arrive at the following formula 
for the average velocity
\begin{equation}
%<3.25>
\bar v=\langle z|\hat v|z\rangle=\frac{c\sigma}{\sqrt{\pi}\hbar}
\int_{-\infty}^\infty\frac{p}{\sqrt{p^2+m^2c^2}}e^{-\frac{\sigma^2}{\hbar^2}(p-
{\bar p})^2}dp.
\end{equation}
Hence, integrating by parts and using (3.14) we obtain
\begin{equation}
%<3.26>
\bar v=\frac{\partial\bar E}{\partial\bar p}.
\end{equation}
Thus, it turns out that $\bar v$ satisfies the classical relation. On the 
other hand, an immediate consequence of (3.16), (3.26) and (A3)
is the power series expansion for $\bar v$ of the form
\begin{equation}
%<3.27>
\bar v=ce^{-\left(\frac{\sigma\bar p}{\hbar}\right)^2}
\sum_{n=0}^{\infty}\frac{1}{n!}\left(\frac{\sigma\bar 
p}{\hbar}\right)^{2n+1}U\left[\frac{1}{2},-n,\left(\frac{\sigma}{\lambda_c}\right)^2\right].
\end{equation}
The numerical calculation based on (3.25) or (3.27) show that
\begin{equation}
%<3.28>                       
\bar v\approx\frac{c\bar p}{\sqrt{{\bar p}^2+m^2c^2}},
\end{equation}
where the approximation is as good as for (3.17) provided $\sigma/\lambda_c\geqslant8$.  
This observation confirms once more that the parameter $\bar p$ marking the coherent states 
can be regarded as the classical relativistic momentum.
\section{Lorentzian coherent states}
\subsection{Lorentzian coherent states}
We now discuss the second candidate for relativistic coherent states 
--- the momentum-space wave functions of the form
\begin{equation}
%<4.1>
\tilde\phi(p)=C\exp(-\alpha\sqrt{p^2+m^2}+\beta p),
\end{equation}
where $\alpha>|{\rm Re}\beta|$, $\hbar=1$, $c=1$, and $C$ is a normalization 
constant, and it is assumed that the Hilbert space is specified by (3.8).  
The wave functions (4.1) for $\beta=0$ were utilized in \cite{25} and \cite{52} for 
constructing the explicit solution to the free Salpeter equation
\begin{equation}
%<4.2>
{\rm i}\frac{\partial\tilde\phi(p,t)}{\partial 
t}=\sqrt{p^2+m^2}\tilde\phi(p,t),
\end{equation}
where $\hbar=c=1$, in the momentum representation.  Namely, we have
\begin{equation}
%<4.3>
\tilde\phi(p,t)=C\exp[-(\alpha+{\rm i}t)\sqrt{p^2+m^2}].
\end{equation}
The denomination 
``Lorentzian'' was introduced by Rosenstein and Horwitz \cite{53} who used it for the wave 
function (4.1) with $m=0$ and $\beta=0$.
On performing the inverse Fourier transformation one can easily obtain the 
solution to the free particle Salpeter equation in the coordinate 
representation.  The conterpart of the wave functions (4.1) in the case of 
the Dirac equation was discussed in \cite{54} in the context of the quantum 
walks. The interpretation of the Lorentzian wave packets (4.1) as 
relativistic coherent states is supported by observations of Al-Hashimi 
and Wiese \cite{55} who considered the general case with $\beta\neq0$ and 
showed that these states minimalize the position-velocity uncertainty relations.  
In spite of this result the arising possibility of interpreting (4.1) as coherent 
states was not analyzed in \cite{55}.  In particular, the closeness of the 
Lorentzian states (4.1) to the classical ones connected with the parametrization 
of the phase space was not investigated therein.

Now, the form of (4.1) indicates that it can be regarded as a relativistic 
generalization of (3.11).  Indeed, applying the Thiemann complexifier 
method \cite{56} we can write the annihilation operator (3.2) as
\begin{equation}
%<4.4>
\hat a=\frac{1}{\sqrt{2}}e^{-\frac{\sigma^2{\hat p}^2}{2\hbar^2}}\frac{\hat x}{\sigma}
e^{\frac{\sigma^2{\hat p}^2}{2\hbar^2}}.
\end{equation}
This operator is the nonrelativistic limit of the following one
\begin{equation}
%<4.5>
\hat b=\frac{1}{\sqrt{2}}e^{-\frac{\sigma^2\sqrt{{\hat p}^2+m^2c^2}}{\lambda_c\hbar}}
\frac{\hat x}{\sigma}e^{\frac{\sigma^2\sqrt{{\hat p}^2+m^2c^2}}{\lambda_c\hbar}},
\end{equation}
where $\lambda_c=\hbar/mc$ is the Compton wavelength.  An immediate 
consequence of (4.5) is
\begin{equation}
%<4.6>
\hat b=\frac{1}{\sqrt{2}}\left(\frac{\hat x}{\sigma}+{\rm i}\frac{\sigma}{\lambda_c}
\frac{\hat p}{\sqrt{{\hat p}^2+m^2c^2}}\right).
\end{equation}
Hence, we finally get
\begin{equation}
%<4.7>
\hat b=\frac{1}{\sqrt{2}}\left(\frac{\hat x}{\sigma}+{\rm i}\frac{\sigma}{\lambda_c}
\frac{\hat v}{c}\right),
\end{equation}
where $\hat v$ is the operator of the relativistic velocity given by 
(3.31).  Now, as with (3.1) we define the coherent states as eigenvectors of the 
operator $\hat b$  
\begin{equation}
%<4.8>
\hat b|\zeta\rangle=\zeta|\zeta\rangle,
\end{equation}
where $\zeta$ is a complex number.  Analogously as in the case of the 
canonical coherent states discussed in the previous section we can express 
the complex number $\zeta$ with the help of average position 
$\bar x~=~\langle\zeta|\hat x|\zeta\rangle$ and average velocity 
$\bar v~=~\langle\zeta|\hat v|\zeta\rangle$, where $|\zeta\rangle$ is a 
normalized coherent state.  Namely, we have
\begin{equation}
%<4.9>
\zeta=\frac{1}{\sqrt{2}}\left(\frac{\bar x}{\sigma}+{\rm i}\frac{\sigma}{\lambda_c}
\frac{\bar v}{c}\right).
\end{equation}
The variations in the coherent state $|\zeta\rangle$ fulfil
\begin{align}
%<4.10>
(\Delta\hat x)^2 &=-{\rm i}\frac{\sigma^2}{2\lambda_cc}\langle\zeta|[\hat 
x,\hat v]|\zeta\rangle,\\
(\Delta\hat v)^2 &=-{\rm i}\frac{c\lambda_c}{2\sigma^2}\langle\zeta|[\hat 
x,\hat v]|\zeta\rangle.
\end{align}
Therefore
\begin{equation}
%<4.12>
(\Delta\hat x)^2(\Delta\hat v)^2=\textstyle{\frac{1}{4}}|\langle\zeta
|[\hat x,\hat v]|\zeta\rangle|^2.
\end{equation}
It thus appears that the states $|\zeta\rangle$ minimalize the Robertson 
position-velocity uncertainty relation
\begin{equation}
%<4.13>
(\Delta\hat x)^2(\Delta\hat v)^2\geqslant\textstyle{\frac{1}{4}}|\langle\zeta
|[\hat x,\hat v]|\zeta\rangle|^2.
\end{equation}
We now return to (4.8).  On writing this equation in the momentum 
representation (see (3.9)) we arrive at the following momentum-space wave function 
representing the coherent state
\begin{equation}
%<4.14>
{\tilde\phi}_\zeta(p)=C\exp\left[-\frac{\sigma^2}{\lambda_c\hbar}\sqrt{p^2+m^2c^2}
-{\rm i}\frac{\sqrt{2}\sigma\zeta}{\hbar}p\right],
\end{equation}
where $C$ is a normalization constant.  Thus, it turns out that the 
abstract coherent states defined by (4.8) coincide with the Lorentzian wave 
functions (4.1) with fixed values of the parameters $\alpha$ and $\beta$ 
specified by (4.14).  We point out that the states (4.1) were derived in \cite{54} in a more
complicated way by demanding minimization of the Heisenberg uncertainty 
relation for position and velocity (4.13).  Using the identity \cite{57}
\begin{align}
%<4.15>
\int_0^\infty\exp(-\mu\sqrt{x^2+\nu^2})\cosh\rho 
x\,dx=&\frac{\mu\nu}{\sqrt{\mu^2-\rho^2}}K_1(\nu\sqrt{\mu^2-\rho^2}),\nonumber\\
&{\rm Re}\mu>|{\rm Re}\rho|,\quad {\rm Re}\nu>0,
\end{align}
where $K_\nu(z)$ is the modified Bessel function (Macdonald function), we find that the 
normalization constant $C$ in (4.14) is given by
\begin{equation}
%<4.16>
C^2=\frac{\lambda_c\sqrt{1+\left(\frac{\lambda_c}{\sigma}\right)^2\left(\frac{\zeta-\zeta^*}
{\sqrt{2}}\right)^2}}{2K_1\left[2\left(\frac{\sigma}{\lambda_c}\right)^2
\sqrt{1+\left(\frac{\lambda_c}{\sigma}\right)^2\left(\frac{\zeta-\zeta^*}{\sqrt{2}}\right)^2}\right]}.
\end{equation}
On the other hand, it appears that we have restriction on the complex number $\zeta$
parametrizing coherent states imposed by the first inequality from (4.15).  
Namely,
\begin{equation}
%<4.17>
|\zeta-\zeta^*|<\sqrt{2}\frac{\sigma}{\lambda_c}.
\end{equation}
Making use of (4.9) we can parametrize the coherent states by $\bar x$ and 
$\bar v$, so
\begin{equation}
%<4.18>
{\tilde\phi}_\zeta(p)=C\exp\left[-\frac{\sigma^2}{\lambda_c\hbar}\sqrt{p^2+m^2c^2}
-{\rm i}\frac{\sigma}{\hbar}\left(\frac{\bar x}{\sigma}+{\rm i}\frac{\sigma}{\lambda_c}
\frac{\bar v}{c}\right)p\right],
\end{equation}
and
\begin{equation}
%<4.19>
C^2=\frac{\lambda_c\sqrt{1-\left(\frac{\bar v}{c}\right)^2}}
{2K_1\left[2\left(\frac{\sigma}{\lambda_c}\right)^2\sqrt{1-\left(\frac{\bar 
v}{c}\right)^2}\right]}.
\end{equation}
The restriction (4.17) on $\zeta$ takes the simple, physically meaningful 
form
\begin{equation}
%<4.20>
\Big\vert\frac{\bar v}{c}\Big\vert<1.
\end{equation}
Now we are in a position to calculate the variances of the position and 
momentum in the coherent states.  Taking into account (4.18) and the 
identity \cite{57}
\begin{align}
%<4.21>
\int_0^\infty\frac{\exp(-\mu\sqrt{x^2+\nu^2})}{\sqrt{x^2+\nu^2}}\cosh\rho 
x\,dx=&K_0(\nu\sqrt{\mu^2-\rho^2}),\nonumber\\
&{\rm Re}\mu>|{\rm Re}\rho|,\quad {\rm Re}\nu>0,
\end{align}
we get
\begin{align}
%<4.22>
|\langle\zeta|[\hat x,\hat v]|\zeta\rangle|
=&\frac{2\sigma^2c}{\lambda_c}\left\{1-\left(\frac{\bar v}{c}\right)^2-
\frac{2\left(\frac{\sigma}{\lambda_c}\right)^2\sqrt{1-\left(\frac{\bar 
v}{c}\right)^2}}{K_1\left[2\left(\frac{\sigma}{\lambda_c}\right)^2\sqrt{1-\left(\frac{\bar 
v}{c}\right)^2}\right]}\right.\nonumber\\
&\times\left.\int_1^{\infty}K_0\left[2\left(\frac{\sigma}{\lambda_c}\right)^2
\sqrt{\xi^2-\left(\frac{\bar v}{c}\right)^2}\right]d\xi\right\}.
\end{align}
Eqs.\ (4.10), (4.11) and (4.22) taken together yield
\begin{align}
%<4.23>
(\Delta\hat x)^2=&\frac{\sigma^4}{\lambda_c^2}
\left\{1-\left(\frac{\bar v}{c}\right)^2-
\frac{2\left(\frac{\sigma}{\lambda_c}\right)^2\sqrt{1-\left(\frac{\bar 
v}{c}\right)^2}}{K_1\left[2\left(\frac{\sigma}{\lambda_c}\right)^2\sqrt{1-\left(\frac{\bar 
v}{c}\right)^2}\right]}\right.\nonumber\\
&\times\left.\int_1^{\infty}K_0\left[2\left(\frac{\sigma}{\lambda_c}\right)^2
\sqrt{\xi^2-\left(\frac{\bar v}{c}\right)^2}\right]d\xi\right\},\\
(\Delta\hat v)^2=& c^2
\left\{1-\left(\frac{\bar v}{c}\right)^2-
\frac{2\left(\frac{\sigma}{\lambda_c}\right)^2\sqrt{1-\left(\frac{\bar 
v}{c}\right)^2}}{K_1\left[2\left(\frac{\sigma}{\lambda_c}\right)^2\sqrt{1-\left(\frac{\bar 
v}{c}\right)^2}\right]}\right.\nonumber\\
&\times\left.\int_1^{\infty}K_0\left[2\left(\frac{\sigma}{\lambda_c}\right)^2
\sqrt{\xi^2-\left(\frac{\bar v}{c}\right)^2}\right]d\xi\right\}.
\end{align}
We point out that in opposition to the case of the canonical coherent 
states (see (3.4)), the variances (4.23) and (4.24) depend on the 
parameter $\bar v$ labelling the Lorentzian coherent states.  The scalar 
product of coherent states calculated with the help of (3.8), (4.14) and 
(4.15) is of the form
\begin{equation}
%<4.25>
\langle\zeta|\eta\rangle=\frac{A(\zeta)A(\eta)}
{\sqrt{1+\left(\frac{\lambda_c}{\sigma}\right)^2\left(\frac{\zeta^*-\eta}
{\sqrt{2}}\right)^2}}K_1\left[2\left(\frac{\sigma}{\lambda_c}\right)^2
\sqrt{1+\left(\frac{\lambda_c}{\sigma}\right)^2\left(\frac{\zeta^*-\eta}
{\sqrt{2}}\right)^2}\right],
\end{equation}
where
\begin{equation}
%<4.26>
A^2(\zeta)=\frac{\sqrt{1+\left(\frac{\lambda_c}{\sigma}\right)^2\left(\frac{\zeta-\zeta^*}
{\sqrt{2}}\right)^2}}{K_1\left[2\left(\frac{\sigma}{\lambda_c}\right)^2
\sqrt{1+\left(\frac{\lambda_c}{\sigma}\right)^2\left(\frac{\zeta-\zeta^*}
{\sqrt{2}}\right)^2}\right]}
\end{equation}
and we have restriction on $\zeta$ and $\eta$ (see (4.15)) such that
\begin{equation}
%<4.27>
\bigg\vert{\rm Re}\left[{\rm i}\frac{\lambda_c}{\sigma}\frac{1}{\sqrt{2}}
(\zeta^*-\eta)\right]\bigg\vert<1
\end{equation}
leading to
\begin{equation}
%<4.28>
\Big\vert\frac{\bar v}{c}+\frac{\bar v'}{c}\Big\vert<2,
\end{equation}
where $\zeta=\frac{1}{\sqrt{2}}\left(\frac{\bar x}{\sigma}+{\rm i}
\frac{\sigma}{\lambda_c}\frac{\bar v}{c}\right)$ and $\eta=\frac{1}{\sqrt{2}}
\left(\frac{\bar x'}{\sigma}+{\rm i}\frac{\sigma}{\lambda_c}\frac{\bar 
v'}{c}\right)$.
Notice that (4.28) is implied by $\big\vert\frac{\bar v}{c}\big\vert<1$, 
$\big\vert\frac{\bar v'}{c}\big\vert<1$ and the triangle inequality.  We 
also point out that non-orthogonality of coherent states is one of their 
most characteristic properties.
\subsection{Averages of relativistic observables in the Lorentzian coherent states}
In order to study the closeness of the discussed Lorentzian coherent 
states to the classical ones we now analyze averages of relativistic 
observables.  We first discuss the expectation value of the momentum.  An 
easy calculation based on (3.8), (3.9), (4.15), (4.18) and elementary 
properties of the Bessel functions gives
\begin{equation}
%<4.29>
\bar p=\langle\zeta|\hat p|\zeta\rangle=\frac{m\bar v}{\sqrt{1-\left(\frac{\bar v}{c}\right)^2}}
\frac{K_2\left[2\left(\frac{\sigma}{\lambda_c}\right)^2\sqrt{1-\left(\frac{\bar v}{c}\right)^2}
\right]}{K_1\left[2\left(\frac{\sigma}{\lambda_c}\right)^2\sqrt{1-\left(\frac{\bar v}{c}\right)^2}\right]}.
\end{equation}
From numerical calculations it follows that for sufficiently large $\sigma/\lambda_c$
\begin{equation}
%<4.30>
\bar p\approx\frac{m\bar v}{\sqrt{1-\left(\frac{\bar v}{c}\right)^2}},
\end{equation}
so we can interpret $\bar v$ labelling the coherent states as the 
classical relativistic velocity.  The approximation (4.30) is as good as in the case of the 
average relativistic velocity in canonical coherent states analyzed in the previous section (see (3.31)).  
Namely (4.30) holds with relative error of order 1\% for $\sigma/\lambda_c\geqslant8$.   
We point out that in opposition to the standard coherent states applied for a relativistic particle
discussed in the previous section, the constant $\sigma$ refers to the nonrelativistic case.  
Nevertheless, taking for example $\sigma~=~8\lambda_c^{(\pi^+)}$, where $\sigma~=~\sqrt{\hbar/m\omega}$ is a 
characteristic oscillator length utilized in the definition of the 
nonrelativistic harmonic oscillator coherent states, and $\lambda_c^{(\pi^+)}$
is the Compton wavelength of pion $\pi^+$, we get $\omega=3.2{\cdot}10^{21}\,\,{\rm Hz}$. 
This value is about one-half of the frequency $\omega_{\rm U}=6.9{\cdot}10^{21}\,\,{\rm Hz}$ 
of the wave-function oscillations corresponding to first orbital of a single electron and nucleus 
of Uranium, so it is physically meaningful.

We now study the average value of the energy in the Lorentzian coherent 
state.  Making use of (3.8), (3.9), (4.18), (4.29) and elementary 
properties of the Bessel functions we find
\begin{align}
%<4.31>
\bar E=\langle\zeta|\sqrt{c^2{\hat p}^2+m^2c^4}|\zeta\rangle=&\frac{mc^2}{\sqrt{1-\left(\frac{\bar v}{c}\right)^2}}
\frac{K_2\left[2\left(\frac{\sigma}{\lambda_c}\right)^2\sqrt{1-\left(\frac{\bar v}{c}\right)^2}
\right]}{K_1\left[2\left(\frac{\sigma}{\lambda_c}\right)^2\sqrt{1-\left(\frac{\bar v}{c}\right)^2}
\right]}\nonumber\\
&-\frac{1}{2}\left(\frac{\lambda_c}{\sigma}\right)^2mc^2.
\end{align}
Numerical calculations show that as with the average momentum, 
whenever $\sigma/\lambda_c\geqslant8$, then we have the approximate formula
\begin{equation}
%<4.32>
\bar E\approx\frac{mc^2}{\sqrt{1-\left(\frac{\bar v}{c}\right)^2}},
\end{equation}
where the approximation is as good as in (4.30).  This observation confirms 
once more that $\bar v$ can be identified with the classical relativistic velocity.

The formulas for the averages (4.29) and (4.31) are most probably new.  An 
equivalent but more complicated form of these relations was obtained in 
\cite{55}, where the parametrization was used by means of the Bessel 
functions $K_0(z)$ and $K_1(z)$ instead of $K_1(z)$ and $K_2(z)$.

We finally remark that the construction of Lorentzian coherent states 
introduced in this section cannot be applied in the case of the massless 
particles.  Indeed, on the one hand, in the case of massless particles 
there is no nonrelativistic limit.  On the other hand, we have no 
classical formulae relating for instance energy or momentum of a massless
particle to its velocity.  Therefore, the application of the Lorentzian 
coherent states labelled by classical velocities instead of momenta 
appears limited to the massive case.  Clearly, the Lorentzian wave 
functions (4.1) can be studied in the limit $m=0$ without reference to the 
theory of coherent states as was done for example in \cite{23} and \cite{55}.
\subsection{Position-momentum uncertainty relations in the Lorentzian coherent states}
Our purpose now is the analysis of the uncertainty relations between 
position and momentum in the Lorentzian coherent states.  Proceeding analogously 
as with (4.29) the following formula for the average value of the squared momentum 
can be easily obtained
\begin{align}
%<4.33>
{\bar p^2}=&\langle\zeta|{\hat p}^2|\zeta\rangle\nonumber\\
&=\left[\frac{m\bar v}{\sqrt{1-\left(\frac{\bar v}{c}\right)^2}}\right]^2+\frac{(mc)^2}{2}
\left(\frac{\lambda_c}{\sigma}\right)^2\frac{1+3\left(\frac{\bar v}{c}\right)^2}
{\left[1-\left(\frac{\bar v}{c}\right)^2\right]^{\frac{3}{2}}}
\frac{K_2\left[2\left(\frac{\sigma}{\lambda_c}\right)^2\sqrt{1-\left(\frac{\bar v}{c}\right)^2}
\right]}{K_1\left[2\left(\frac{\sigma}{\lambda_c}\right)^2\sqrt{1-\left(\frac{\bar 
v}{c}\right)^2}\right]}.
\end{align} 
Eqs.\ (4.29) and (4.33) taking together yield
\begin{align}
%<4.34>
(\Delta\hat p)^2=&\left[\frac{m\bar v}{\sqrt{1-\left(\frac{\bar 
v}{c}\right)^2}}\right]^2\left[1-\left(
\frac{K_2\left[2\left(\frac{\sigma}{\lambda_c}\right)^2\sqrt{1-\left(\frac{\bar v}{c}\right)^2}
\right]}{K_1\left[2\left(\frac{\sigma}{\lambda_c}\right)^2\sqrt{1-\left(\frac{\bar v}{c}\right)^2}\right]}
\right)^2\right]\nonumber\\
&{}+\frac{(mc)^2}{2}\left(\frac{\lambda_c}{\sigma}\right)^2\frac{1+3\left(\frac{\bar v}{c}\right)^2}
{\left[1-\left(\frac{\bar v}{c}\right)^2\right]^{\frac{3}{2}}}
\frac{K_2\left[2\left(\frac{\sigma}{\lambda_c}\right)^2\sqrt{1-\left(\frac{\bar v}{c}\right)^2}
\right]}{K_1\left[2\left(\frac{\sigma}{\lambda_c}\right)^2\sqrt{1-\left(\frac{\bar 
v}{c}\right)^2}\right]}.
\end{align} 
The formulas (4.29) and (4.31) are new.  Their more complicated equivalents 
expressed with the help of the Bessel functions $K_0(z)$ and $K_1(z)$ were introduced in 
\cite{22}.

The plots of the rescaled dimensionless variances $(\Delta{\hat x}/\sigma)^2$, 
$(\Delta\sigma{\hat p}/\hbar)^2$, and their product versus $\bar v/c$ are illustrated in Fig. 
1,2 and 3, respectively, where we have chosen the scale according
to (3.4), so that for the canonical coherent states the variances are 1/2 and their 
product is 1/4.  We point out that while we have the better localization of the position 
observable described by its variance than in the case of the canonical coherent states, the
variance of the momentum observable and product of the variances is growing with $\bar v/c$.  
\begin{figure*}
\centering
\includegraphics[scale=1]{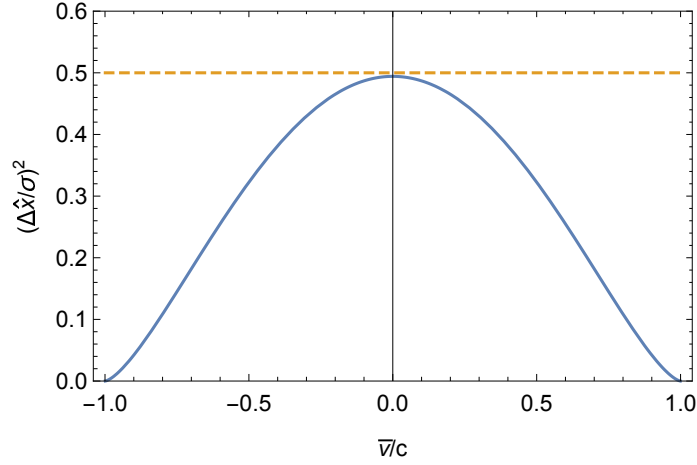}
\caption{Plot of the variance of the position in the Lorentzian coherent state 
given by (4.23), where $\sigma/\lambda_c=8$.  The dashed line corresponds to the value 
$0.5$ of the variance in the canonical coherent state (see (3.4)).}
\end{figure*}
\begin{figure*}
\centering
\includegraphics[scale=1]{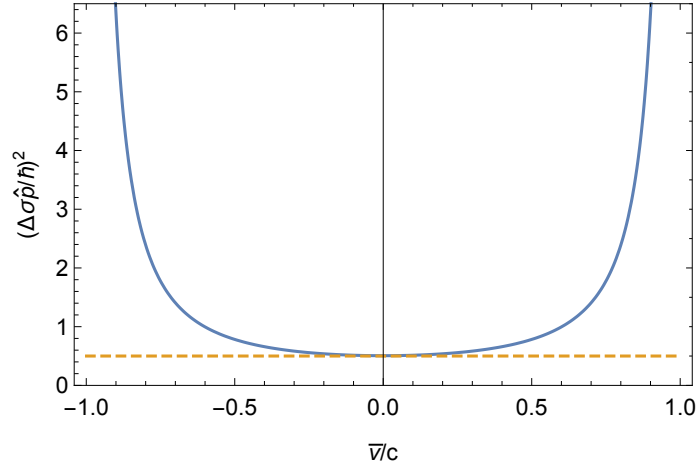}
\caption{Variance of the momentum in the Lorentzian coherent state (4.34)
given by (4.23) with $\sigma/\lambda_c=8$.  The dashed line refers to the variance 
$0.5$ in the canonical coherent state (see (3.4)).}
\end{figure*}
\begin{figure*}
\centering
\includegraphics[scale=1]{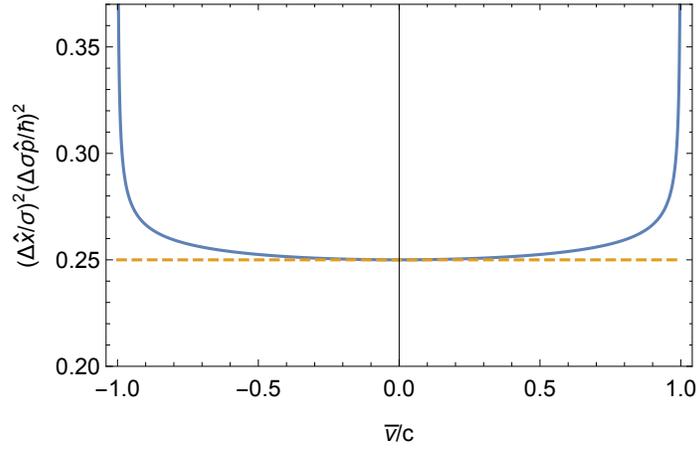}
\caption{Plot of the product of variances of the position and momentum illustrated 
in Fig.\ 1 and Fig.\ 2.  The dashed line corresponds to the value $0.25$ of the product 
in the canonical coherent state (see (3.4)).}
\end{figure*}
\section{Poincar\'e coherent states}
\subsection{Basic properties of the Poincar\'e coherent states}
We now study the candidate for the relativistic coherent states introduced by 
Kaiser \cite{43} and independently by Twareque Ali, Antoine and Gazeau \cite{44,45}. 
The special case of these states was discussed for the first time by Bakke and 
Wergeland \cite{41,42}.  More precisely, the states considered by Bakke and 
Wergeland as packets that do not violate relativistic causality are of the form 
\cite{42}
\begin{equation}
%<5.1>
\tilde\phi(p)=C\frac{\exp(-\alpha\sqrt{p^2+m^2})}{(p^2+m^2)^{\frac{1}{4}}},
\end{equation}
where $\alpha$ is positive parameter, $C$ is a normalization constant, 
$\hbar=1$, $c=1$, and we assume that the Hilbert space is given by (3.8).
We point out that the states (5.1) were not interpreted by Baake and 
Wergeland as the coherent ones.  By virtue of (2.9) these states are equivalent 
in the one-dimensional counterpart of the Hilbert space (2.8) such that
\begin{equation}
%<5.2>
\langle\phi'|\psi'\rangle = \frac{1}{\hbar}\int\frac{mc\,dp}{p^0(p)}
{\tilde\phi}'^*(p){\tilde\psi}'(p),
\end{equation}
to the state
\begin{equation}
%<5.3>
\tilde\phi'(p)=C\exp(-\alpha\sqrt{p^2+m^2}).
\end{equation}

The most general and systematic definition of the relativistic coherent 
states that can be regarded as a generalization of the states (5.3) was 
introduced by Twareque Ali, Antoine and Gazeau \cite{44,45}.  Bearing in 
mind the denomination of these states utilized in \cite{44,45} we shall
refer to them as the Poincar\'e coherent states.  Based on the Perelomov 
approach \cite{58} the above-mentioned coherent states are defined as
\begin{equation}
%<5.4>
|q,k\rangle = U[\sigma_\tau(q,k)]|\psi_0\rangle,
\end{equation}
where $U$ is an element of the unitary irreducible representation of the 
Poincar\'e group ${\cal P}^\uparrow_+(1,1)$ associated to elementary quantum 
system of mass $m>0$, $\sigma_\tau:\Gamma\to{\cal P}^\uparrow_+(1,1)$, 
is a section given by (B3) and $\Gamma={\cal P}^\uparrow_+(1,1)/T_t$, where $T_t$ is the 
subgroup of time translations, is the classical phase space with coordinates 
$(q,k)$. The fiducial vector $|\psi_0\rangle$ is referred to in \cite{45} as 
a ``probe''.  On writing (5.4) in the momentum representation with invariant measure 
(5.2), we get the following formula for the coherent states
\begin{equation}
%<5.5>
\tilde\phi'_{(q,k)}(p)=U(\sigma_\tau(q,k))\psi_0(p)=\exp\left\{{\rm i}
\left[\frac{k_0}{m}p_0\tau-\left(\frac{k}{m}p\tau+pq\right)\right]\right\}
\psi_0(\Lambda_k^{-1}p),
\end{equation}
where we set $\hbar=1$.  Now, it is clear that in the discussed 
one-dimensional case the phase space should be parametrized solely by $q$ 
and $k$, so we set $\tau=0$.  We remark that the section $\sigma_0(q,k)$ 
is called in \cite{45} ``Galilean'' or ``time zero'' section.  Furthermore, 
following \cite{45} we put $\psi_0(p)=Ce^{-\frac{p_0}{\kappa}}$, where $C$ 
is a normalization constant and $\kappa$ is a constant with the dimension 
of energy.  Hence, we finally get
\begin{equation}
%<5.6>
\tilde\phi'_{(q,k)}(p)=U(\sigma_0(q,k))Ce^{-\frac{p_0}{\kappa}}=
C\exp\left(-\frac{1}{\kappa m}k_0p_0+\frac{1}{\kappa m}kp-{\rm i}pq\right),
\end{equation}
where the normalization constant $C$ obtained by means of (5.2) and (4.21) is
\begin{equation}
%<5.7>
C=\sqrt{\frac{\hbar}{2mcK_0\left(\frac{2mc^2}{\kappa}\right)}}.
\end{equation} 
The coherent states (5.6) can be written as
\begin{equation}
%<5.8>
\tilde\phi'_{(q,k)}(p)=Ce^{{\rm i}\underbar{\hbox{$\scriptstyle p$}}\,\underbar{\hbox{$\scriptstyle z$}}},
\end{equation}
where $\underbar{$p$}\,\underbar{$z$}=p_0z_0-pz$, $\underbar{$p$}$ is the 
four-momentum $\underbar{$p$}=(p_0,p)$ with $p_0(p)=\sqrt{p^2+m^2c^2}$, 
and $\underbar{$z$}$ is the complex four-vector $\underbar{$z$}=(z_0,z)=\left({\rm i}\frac{k_0}
{\kappa m},q+{\rm i}\frac{k}{\kappa m}\right)$.

We remark that it is straightforward to extend the construction of 
Poincar\'e coherent states described above to higher dimensions.  The 
coherent states introduced by Kaiser \cite{43} can be considered as such a 
generalization. More precisely, these states were defined in the Hilbert space 
with Lorentz invariant measure (5.2) as
\begin{equation}
%<5.9>
\tilde\phi'_z(p)=Ce^{{\rm i}\underbar{\hbox{$\scriptstyle p$}}\,\underbar{\hbox{$\scriptstyle z$}}},
\end{equation}
where $\underbar{$p$}\,\underbar{$z$}=p_0z_0-{\bm z}\cdot{\bm p}$, $\underbar{$p$}$ is the 
four-momentum: $\underbar{$p$}=(p_0,{\bm p})$, with $p_0({\bm p})=\sqrt{{\bm p}^2+m^2c^2}$,
$\underbar{$z$}$ is a complex four-vector such that $\underbar{$z$}=(a_0+{\rm i}b_0,{\bm a}+{\rm i}{\bm b})$,
and $a_0$ is identified with $-ct$, so $\tilde\phi'_z(p)$ are the time-dependend
coherent states, where the dynamics is given by the Salpeter equation. The coherent state (5.9) 
has the finite norm only if $b_0^2-{\bm b}^2>0$, therefore we introduce the timelike four-vector 
$(b_0,{\bm b})$ satisfying $b_0^2-{\bm b}^2=\eta^2$, where $\eta>0$.   Hence, we obtain
\begin{equation}
%<5.10>
\tilde\phi'_z({\bm p})=C\exp[-\sqrt{{\bm b}^2+\eta^2}\sqrt{{\bm p}^2+m^2c^2}+
{\bm b}\cdot{\bm p}+{\rm i}(-ct\sqrt{{\bm p}^2+m^2c^2}-{\bm a}\cdot{\bm p})].
\end{equation} 
We point out that the generalization of (5.9) was discussed in \cite{43} 
referring to the case, when the vectors are elements of $n+1$-dimensional Minkowski
space.  Now, taking into account that the definition of coherent states should not 
depend on dynamics but a parametrization of a classical pase space, we set 
$t=0$.  We also restrict for simplicity to the one dimensional case, so
the Hilbert space of states is then given by (5.2). With these assumptions
the coherent states (5.10) take the form
\begin{equation}
%<5.11>
\tilde\phi'_z(p)=C\exp(-\sqrt{b^2+\eta^2}\sqrt{p^2+m^2c^2}+bp-{\rm i}ap),
\end{equation} 
where the normalization constant $C$ obtained with the help of (5.2) and
(4.21) is given by
\begin{equation}
%<5.12>
C^2=\frac{\hbar}{2mcK_0(2mc\sqrt{b_0^2-b^2})}=\frac{\hbar}{2mcK_0(2mc\eta)}.
\end{equation}
Comparing (5.6) and (5.7) with (5.11) and (5.12), respectively,
we find that (5.6) and (5.11) represent the same coherent state. In the
sequel we shall adopt parametrization of the coherent states utilized in
(5.11).

We now analyze the physical meaning of the discussed coherent states.
Using (5.2) we get the expectation value of the 
one-dimensional version of the Newton-Wigner (2.10) of the form
\begin{equation}
%<5.13>
{\hat x}_{\rm NW}={\rm i}\hbar\frac{\partial}{\partial p}
-{\rm i}\frac{\hbar}{2}\frac{p}{p^2+m^2c^2},
\end{equation}
in the coherent state (5.11).  Namely, we have
\begin{equation}
%<5.14>
{\bar x}=a\hbar.
\end{equation}
Furthermore, utilizing (5.2) and (4.15) we get the average value 
of the momentum in the Poincar\'e coherent state
\begin{equation}
%<5.15>
\bar p=\frac{bmc}{\sqrt{b_0^2-b^2}}\frac{K_1(2mc\sqrt{b_0^2-b^2})}{K_0(2mc\sqrt{b_0^2-b^2})}
=\frac{bmc}{\eta}\frac{K_1(2mc\eta)}{K_0(2mc\eta)}.
\end{equation}
Taking into account the nonrelativistic limit of the Poincar\'e coherent 
states (compare (4.18)), we get
\begin{equation}
%<5.16>
\qquad\eta=\frac{\sigma^2}{\lambda_c\hbar}.
\end{equation}
Hence, we have
\begin{equation}
%<5.17>
b=\frac{\sigma^2}{\hbar^2}\bar p\rho\left(\frac{\sigma}{\lambda_c}\right),\quad
b_0=\sqrt{\eta^2+b^2}=\frac{\sigma^2}{\hbar^2}\sqrt{m^2c^2+{\bar p}^2
\left[\rho\left(\frac{\sigma}{\lambda_c}\right)\right]^2},
\end{equation}
where we set for brevity 
$\rho\left(\frac{\sigma}{\lambda_c}\right)=K_0\left[2\left(\frac{\sigma}{\lambda_c}\right)^2\right]/
K_1\left[2\left(\frac{\sigma}{\lambda_c}\right)^2\right]$, and we finally obtain the Poincar\'e 
coherent states parametrized by the points of the classical phase space such that
\begin{align}
%<5.18>
\tilde\phi'_{(\bar x,\bar p)}(p)=&\sqrt{\frac{\hbar}{2mcK_0\left[2\left(\frac{\sigma}{\lambda_c}\right)^2\right]}}
\exp\left(-\frac{\sigma^2}{\hbar^2}\sqrt{m^2c^2+{\bar p}^2
\left[\rho\left(\frac{\sigma}{\lambda_c}\right)\right]^2}\sqrt{p^2+m^2c^2}\right.\nonumber\\
&\left.{}+\frac{\sigma^2}{\hbar^2}{\bar p}\rho\left(\frac{\sigma}{\lambda_c}\right)p-
{\rm i}\frac{{\bar x}}{\hbar}p\right).
\end{align}
where $\tilde\phi'_{(\bar x,\bar p)}(p)\equiv\tilde\phi'_{(q(\bar x,\bar p),k(\bar x,\bar p))}(p)
\equiv\tilde\phi'_{z(\bar x,\bar p)}(p)$.
We stress that neither the physical meaning of the invariant $\eta$ nor 
the parametrization of the coherent states by points of the classical phase 
space were discussed in \cite{43}.  We point out that for $\bar p=0$ and ${\bar x}=0$
the states (5.18) reduce to the wave functions (5.3) introduced by Bakke 
and Wergeland.  We also remark that in opposition to the canonical coherent 
states and the Lorentzian ones, the Poincar\'e coherent states are not 
marked by a complex number parametrizing the points of the classical phase 
space $(\bar x,\bar p)$.  Furthermore, the reason is unclear for calling 
the states (5.18) the coherent ones.  For instance their interpretation is 
dim as the states closest to the classical ones. 
\subsection{Position-momentum uncertainty relations in the Poincar\'e coherent states}
We now discuss the Heisenberg position-momentum uncertainty relations in 
the Poincar\'e coherent states.  The variance of the position obtained with 
the use of the completeness of the momentum eigenvectors $|p\rangle$ spanning 
the representation (5.2) can be written as
\begin{align}
%<5.19>
&(\Delta{\hat x}_{\rm NW})^2 =\overline{({\hat x}_{\rm NW}-{\bar x})^2}
=\frac{\hbar^2}{2K_0\left[2\left(\frac{\sigma}{\lambda_c}\right)^2\right]}
\int\frac{dp}{\sqrt{p^2+m^2c^2}}\nonumber\\
&\times\left(-\frac{\sigma^2}{\hbar^2}\sqrt{m^2c^2+{\bar p}^2
\left[\rho\left(\frac{\sigma}{\lambda_c}\right)\right]^2}\frac{p}
{\sqrt{p^2+m^2c^2}}
+\frac{\sigma^2}{\hbar^2}{\bar p}\rho\left(\frac{\sigma}{\lambda_c}\right)-
\frac{1}{2}\frac{p}{p^2+m^2c^2}\right)^2\nonumber\\
&\times\exp\left(-2\frac{\sigma^2}{\hbar^2}\sqrt{m^2c^2+{\bar p}^2
\left[\rho\left(\frac{\sigma}{\lambda_c}\right)\right]^2}\sqrt{p^2+m^2c^2}
+2\frac{\sigma^2}{\hbar^2}{\bar p}\rho\left(\frac{\sigma}{\lambda_c}\right)p\right).
\end{align}
We do not know any other formula for the variance simpler than the integral (5.19).  
Using (4.21) and elementary properties of the Bessel functions we find after some calculation 
\begin{equation}
%<5.20>
(\Delta\hat p)^2=\frac{1}{2}\frac{\hbar^2}{\sigma^2}\frac{1}{\rho\left(\frac{\sigma}{\lambda_c}\right)}
+{\bar p}^2
\left\{\rho\left(\frac{\sigma}{\lambda_c}\right)\frac{K_2\left[2\left(\frac{\sigma}{\lambda_c}\right)^2\right]}
{K_1\left[2\left(\frac{\sigma}{\lambda_c}\right)^2\right]} 
-1\right\}.
\end{equation}
The dimensionless variances $(\Delta{\hat x}_{\rm NW}/\sigma)^2$, 
$(\Delta\sigma{\hat p}/\hbar)^2$, and their product are depicted in Fig. 
4,5 and 6, respectively. As with the case of the Lorentzian coherent 
states we deal with better localization accuracy of position observable in 
comparison with the canonical coherent states and increasing of the 
variance of the momentum observable.  Nevertheless, in opposition to the 
Lorentzian coherent states for the Poincar\'e ones we have plateau on the 
plot of the product of variances of the position and momentum observables.
\begin{figure*}
\centering
\includegraphics[scale=1]{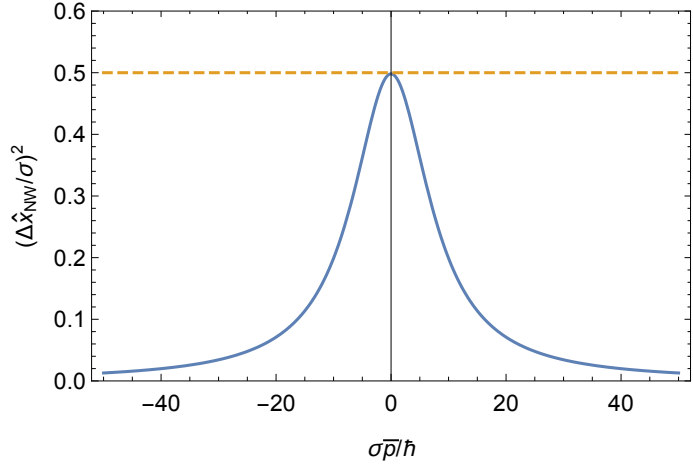}
\caption{Variance of the position in the Poincar\'e coherent state (5.18)
specified by (5.19) with $\sigma/\lambda_c=8$.  The dashed line corresponds to the value 
$0.5$ of the variance in the canonical coherent state (see (3.4)).}
\end{figure*}
\begin{figure*}
\centering
\includegraphics[scale=1]{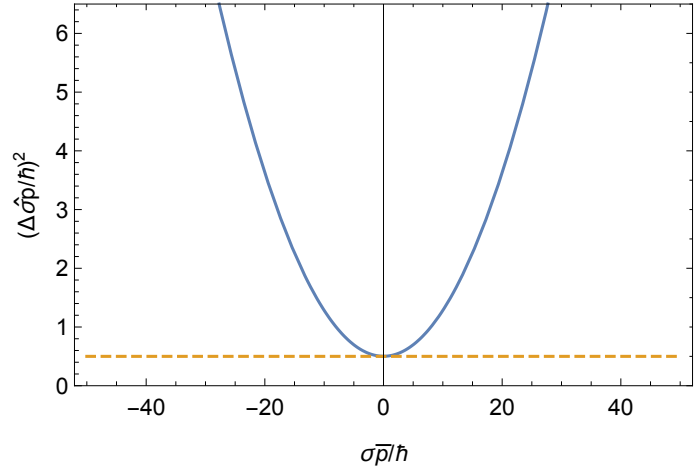}
\caption{Plot of the variance of the momentum in the Poincar\'e coherent state (5.18)
given by (5.20), where $\sigma/\lambda_c=8$.  The dashed line refers to the variance 
$0.5$ in the canonical coherent state (see (3.4)).}
\end{figure*}
\begin{figure*}
\centering
\includegraphics[scale=1]{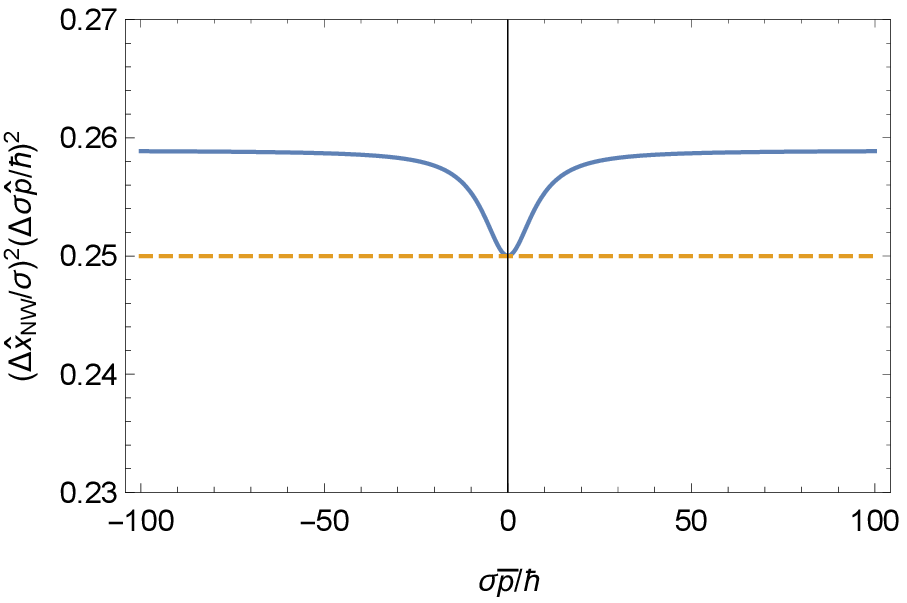}
\caption{The product of variances of the position and momentum illustrated 
in Fig.\ 4 and Fig.\ 5.  The dashed line corresponds to the value $0.25$ of the product 
in the canonical coherent state (see (3.4)).}
\end{figure*}

The Poincar\'e coherent states are not orthogonal.  On making use of the 
completeness of eigenvectors $|p\rangle$ of the momentum operator we 
arrive at the following formula for the overlapping integral
\begin{align}
%<5.21>
&\langle\bar x,\bar p|{\bar x}',{\bar p}'\rangle\nonumber
=\frac{1}{K_0\left[2\left(\frac{\sigma}{\lambda_c}\right)^2\right]}
K_0\left\{mc\left[\left(\frac{\sigma}{\hbar}\right)^4\left(\sqrt{m^2c^2+{\bar p}^2
\left[\rho\left(\frac{\sigma}{\lambda_c}\right)\right]^2}\right.\right.\right.\nonumber\\
&\left.\left.\left.{}+\sqrt{m^2c^2+{\bar p}'^2
\left[\rho\left(\frac{\sigma}{\lambda_c}\right)\right]^2}\right)^2
-\left(\frac{\sigma^2}{\hbar^2}\rho\left(\frac{\sigma}{\lambda_c}\right)({\bar p}+{\bar p}')+{\rm i}
\frac{{\bar x}-{\bar x}'}{\hbar}\right)^2\right]^\frac{1}{2}\right\},
\end{align}
where $\tilde\phi'_z(p)\equiv\langle p|{\bar x},{\bar p}\rangle$.
The Poincar\'e coherent states form a complete set.  Indeed, using the 
orthogonality condition satisfied by the momentum eigenvectors spanning
the basis of the representation (5.2) such that
\begin{equation}
%<5.22>
\langle p|p'\rangle = \frac{\hbar}{mc}\sqrt{p^2+m^2c^2}\delta(p-p'),
\end{equation}
we get the resolution of the identity for the Poincar\'e coherent states of 
the form
\begin{equation}
%<5.23>
\frac{1}{2\pi\hbar}\left[\rho\left(\frac{\sigma}{\lambda_c}\right)\right]^2
\int d{\bar x} d\bar p\,|\bar x,\bar p\rangle
\langle{\bar x},\bar p|=I. 
\end{equation}
It thus appears that up to multiplicative constant the resolution of the
identity for the Poincar\'e coherent states has the same form as in
the case of the canonical states (see (3.7)). 
\subsection{Expectations of relativistic observables in the Poincar\'e coherent states}
We now investigate the behavior of relativistic observables in the Poincar\'e 
coherent states that can be regarded as a measure of the closeness of these states to the 
classical phase space. Taking into account (4.15) we derive the following formula for the 
expectation value of the energy 
\begin{equation}
%<5.24>
\bar E=\overline{\sqrt{{\hat p}^2c^2+m^2c^2}}=\sqrt{{\bar p}^2c^2+
\left[\frac{mc^2}{\rho\left(\frac{\sigma}{\lambda_c}\right)}\right]^2}.
\end{equation}
Notice that as with the Lorentzian coherent states, the average energy
and momentum do not satisfy the relativistic dispersion relation. We also 
remark that
\begin{equation}
%<5.25>
\frac{1}{c}\sqrt{\left(\frac{{\bar E}}{c}\right)^2-{\bar p}^2}=
\frac{m}{\rho\left(\frac{\sigma}{\lambda_c}\right)}=
m\frac{K_1\left[2\left(\frac{\sigma}{\lambda_c}\right)^2\right]}
{K_0\left[2\left(\frac{\sigma}{\lambda_c}\right)^2\right]},
\end{equation}
was interpreted in \cite{43} as the ``effective" mass of a particle, with
the parameter $K_1\left[2\left(\frac{\sigma}{\lambda_c}\right)^2\right]/
K_0\left[2\left(\frac{\sigma}{\lambda_c}\right)^2\right]$ playing the role 
of a renormalization which takes into effect the fluctuation in 
energy-momentum.  Furthermore, the average velocity in the Poincar\'e coherent 
state that can be easily obtained with the help of (4.15) is given by
\begin{equation}
%<5.26>
\bar v = \overline{\left(\frac{c\hat p}{\sqrt{{\hat p}^2+m^2c^2}}\right)}=
\frac{2\left(\frac{\sigma}{\lambda_c}\right)^2\frac{\bar p}{m}}
{K_1\left[2\left(\frac{\sigma}{\lambda_c}\right)^2\right]}
\int_{\frac{\sigma^2}{\lambda_c\hbar}}^\infty\frac{K_1(2mcu)}
{\sqrt{u^2+\left(\frac{\sigma}{\hbar}\right)^4{\bar p}^2
\left[\rho\left(\frac{\sigma}{\lambda_c}\right)\right]^2}}du.
\end{equation}
Numerical calculations show that for $\sigma/\lambda_c\geqslant5$ and
$\sigma/\lambda_c\geqslant8$ we have the approximate relations of the form
(3.19) and (3.31), repectively, so the behavior of expectation values of
energy and velocity in the Poincar\'e coherent states is the same as in the 
case of the canonical ones.
\section{Conclusions}
In this work we discuss the three approaches to relativistic coherent states of a 
spinless particle: the canonical coherent states, Lorentzian and 
Poincar\'e ones. An advantage of the canonical coherent states is saturation of 
the Heisenberg uncertainty relations and possibility of their application 
in the case of the massless particles.  Furthermore, the behavior of 
expectation values of quantum observables in these states is as good or 
better than for the remaining candidates.  The problems are connected with 
the Lorentz covariance of the canonical coherent states.  An asset of the 
Lorentzian coherent states is the minimization of uncertainty relations 
involving position and relativistic velocity and analytic form of averages 
of observables expressed in terms of the Bessel functions that are 
ubiquitous in relativistic quantum theory.  Furthermore, these states can be 
obtained by means of the Thiemann complexifier method that has already proven 
its generality in the case of the canonical coherent states and coherent states 
for a particle on a circle and sphere \cite{59}.  The problematic issue are 
position-momentum uncertainty relations.  More precisely, the behavior of the
variance of the momentum observable and product of variances of position and 
momentum operators that is unusual for coherent states.  Another problem is 
the completeness of the Lorentzian coherent states.  Indeed, the formula for 
the resolution of the identity for these states should involve integral 
identities satisfied by the Bessel functions on the interval  
$(-1,1)$.  The authors have not found such identities in 
the literature.  Finally, it is not clear why in the relativistic quantum
mechanics the coherent states would be parametrized by average position $\bar x$
and velocity $\bar v$ instead of the point of the classical phase space 
$(\bar x,\bar p)$ as in the nonrelativistic case. An advantage of the 
Poincar\'e coherent states is their Lorentz covariance.  Furthermore, the 
plateau on the plot of the product of variances of position and momentum 
variables resembles the behavior of canonical coherent states.  The 
parametrization of the Poincar\'e coherent states by means of the Bessel 
functions seems to be also plausible in relativistic quantum mechanics.
As with the canonical coherent states these states are parametrized by 
points $(\bar x,\bar p)$ of the classical phase space.  We have also 
similar resolution of the identity.  On the other hand, the behavior is 
problematic of the variance of the momentum observable analogous to that in 
the Lorentzian coherent states.  In the light of the above comments, there 
is no convincing evidence which candidate for the relativistic coherent states 
is better.
\section*{Acknowledgements}
This work has been supported by the Polish National Science Centre under Contract
2014/15/B/ST2/00117 and by the University of Lodz.
\appendix
\section{Confluent hypergeometric function}
We first recall some properties of the confluent hypergeometric function
$U(a,b,z)$.  This function can be expressed by means of the hypergeometric
function ${}_1F_1(a,b;z)$.  Namely,
\begin{equation}
%<a1>
U(a,b,z)=\frac{\Gamma(1-b)}{\Gamma(a+1-b)}{}_1F_1(a,b;z)+\frac{\Gamma(b-1)}{\Gamma(a)}
z^{1-b}{}_1F_1(a+1-b,2-b;z).
\end{equation}
The confluent hypergeometric function satisfies the following relations
\begin{align}
%<a2>
U(a,b,z) &=z^{1-b}U(a-b+1,2-b,z),\\
\frac{d^n}{dz^n}[z^{b-1}U(a,b,z)] &=(-1)^n(a-b+1)_n z^{b-n-1}U(a,b-n,z),
\end{align}
where $(\alpha)_n$ is the Pochhammer's symbol 
$(\alpha)_n=\alpha(\alpha+1)\ldots (\alpha+n-1)$ and $(\alpha)_0=1$.
We have also the recurrence relation fulfilled by this function such that
\begin{equation}
%<a4>
U(a,b,z)-aU(a+1,b,z)-U(a,b-1,z)=0.
\end{equation}
\section{Generation of the Poincar\'e coherent states}
We now discuss the details of the generation of the Poincar\'e
coherent states described by the formula (5.4).  In order to make 
the formalism introduced in \cite{8} more transparent we use the 
following representation of the Poincar\'e group 
${\cal P}^\uparrow_+(1,1)$ element $g(\Lambda,\underbar{a})$ with $\underbar{a}=
\left(\begin{smallmatrix}a^0\\a^1\end{smallmatrix}\right)$:
\begin{equation}
%<5.5>
g(\Lambda,\underbar{a}) = 
\begin{pmatrix}
\hphantom{0}\Lambda\hphantom{0}&\underbar{a}\\
0\hphantom{0}0&1
\end{pmatrix}
\end{equation}
Furthermore, the parametrization of the phase space $\Gamma={\cal P}^\uparrow_+(1,1)/T_t$
utilizes the following factorization of the element $g(\Lambda_k,\underbar{x})$, where 
$\Lambda_k=\frac{1}{m}\left(\begin{smallmatrix}k^0&k\\k&k^0\end{smallmatrix}\right)$
is the Lorentz boost matrix and $\underbar{x}=\left(\begin{smallmatrix}t\\x\end{smallmatrix}\right)$:
\begin{equation}
%<5.6>
g(\Lambda_k,\underbar{x}) = 
\begin{pmatrix}
\hphantom{0}\Lambda_k\hphantom{0}&\underbar{x}\\
0\hphantom{0}0&1
\end{pmatrix}=
\begin{pmatrix}
\hphantom{0}\Lambda_k\hphantom{0}&\begin{pmatrix}0\\
q
\end{pmatrix}\\
0\hphantom{0}0&1
\end{pmatrix}
\begin{pmatrix}
\hphantom{0}I\hphantom{0}&\begin{pmatrix}\tau\\
0
\end{pmatrix}\\
0\hphantom{0}0&1
\end{pmatrix}
\end{equation}
which leads to the section of the form
\begin{equation}
%<5.7>
\sigma_\tau(q,k)=
\begin{pmatrix}
\hphantom{0}\Lambda_k\hphantom{0}&\frac{1}{m}\begin{pmatrix}\tau k^0\\
\tau k+mq
\end{pmatrix}\\
0\hphantom{0}0&1
\end{pmatrix}.
\end{equation}


\begin{thebibliography}{}
\bibitem{1}E. Schr\"odinger, Ann. Phys. 81 (1926) 109.
\bibitem{2}O. Klein, Z. Phys. 37 (1926) 895.
\bibitem{3}V.A. Fock, Z. Phys. 38 (1926) 242; V.A. Fock, Z. Phys. 39 (1926) 226.
\bibitem{4}W. Gordon, Z. Phys. 40 (1926) 117; W. Gordon, Z. Phys. 40 (1926) 121.
\bibitem{5}E.E. Salpeter, Phys. Rev. 87 (1952) 328.
\bibitem{6}L.L. Foldy, Phys. Rev. 102 (1956) 568.
\bibitem{7}G. Paiano, Nuovo Cimento 70 (1982) 339.
\bibitem{8}P. Cea, P. Calangelo, G. Nardulli, G. Paiano and G.
Preparata, Phys. Rev. D 26 (1982) 1157.
\bibitem{9}P. Cea, G. Nardulli and G. Paiano, Phys. Rev. D 28 (1983) 2291.
\bibitem{10}L.M. Nickisch and L. Durand, Phys. Rev D 30 (1984) 660.
\bibitem{11}J.B. Rosenstein and L.P. Horwitz, J. Phys. A 18 (1985) 2115.
\bibitem{12}C. L\"ammerzahl, J. Math. Phys. 34 (1993) 3918.
\bibitem{13}W. Lucha and F.F. Sch\"oberl, Phys. Rev. D 50 (1994) 5443.
\bibitem{14}F. Brau, J. Math. Phys. {\bf 39}, 2254 (1998).
\bibitem{15}R.L. Hall, W. Lucha and F.F. Sch\"oberl, J. Phys. A 34 (2001) 5059.
\bibitem{16}R.L. Hall, W. Lucha and F.F. Sch\"oberl, J. Math. Phys. 42 (2001) 5228.
\bibitem{17}R.L. Hall, W. Lucha and F.F. Sch\"oberl, J. Math. Phys.43 (2002) 5913.
\bibitem{18}R.L. Hall, W. Lucha and F.F. Sch\"oberl, Int. J. Mod. Phys. A 18 (2003) 2657.
\bibitem{19}F. Brau, Phys. Lett. A 313 (2003) 363.
\bibitem{20}Zhi-Feng Li, Jin-Jin Liu, W. Lucha and F.F. Sch\"oberl, J. Math. Phys. 46 (2005) 103514.
\bibitem{21}R.L. Hall and W. Lucha, J. Phys. A 38 (2005) 7997.
\bibitem{22}Y. Chargui, L. Chetouani and A. Trabelsi, J. Phys. A: Math. Theor. 42 (2009) 355203.
\bibitem{23}K. Kowalski and J. Rembieli\'nski, Phys. Rev. A 81 (2010) 012118.
\bibitem{24}D. Babusci, G. Dattoli and M. Quattromini, Phus. Rev. A 83 (2011) 062109.
\bibitem{25}K. Kowalski and J. Rembieli\'nski, Phys. Rev. A 84 (2011) 012108.
\bibitem{26}P. Garbaczewski and V. Stephanovich, J. Math. Phys. 54 (2013) 072103.
\bibitem{27}M. {\.Z}aba and P. Garbaczewski, J. Math. Phys. 55 (2014) 092103.
\bibitem{28}G. Dattoli, E. Sabia, K. G\'orska, A. Horzela and K. A. Penson, J. Phys. A: Math. Theor. 48 (2015) 125203.
\bibitem{29}F. Gross, Relativistic Quantum Mechanics and Field Theory, Wiley, Weinheim, 1993.
\bibitem{30}J. Dimock, Quantum Mechanics and Quantum Field Theory. 
A Mathematical Primer, Cambridge University Press, Cambridge, 2011.
\bibitem{31}K. Kowalski and J. Rembieli\'nski, Ann. Phys.  375 (2016) 1. 
\bibitem{32}O.I. Zavialov and A.M. Malokostov, Theor. Math. Phys. 119 (1999) 448.
\bibitem{33}L.J. Nickisch and L. Durand, Phys. Rev. D  30 (1984) 660.
\bibitem{34}J.L. Basdevant and S. Boukraa, Z. Phys. C 28 (1985) 413.
\bibitem{35}I.A. Malkin and V.I. Man'ko, Sov. Phys. JETP 28 (1969) 527.
\bibitem{36}B.I. Lev, A.A. Semenov, C.V. Usenko and J.R. Klauder, Phys. Rev. A 66 (2002) 022115.
\bibitem{37}B.I. Lev, A.A. Semenov and C.V. Usenko, Phys. Lett. A 230 (1997) 261.
\bibitem{38}V.G. Bagrov, I.L. Buchbinder and D.M. Gitman, J. Phys. A: Math. Gen. 9 (1976) 1955.
\bibitem{39}H. Feshbach and F. Villars, Rev. Mod. Phys. 30 (1958) 24.
\bibitem{40}A. Mostafazadeh and F. Zamani, Ann. Phys. 321 (2006) 2210.
\bibitem{41}F. Baake and H. Wergeland, Physica 69 (1973) 5.
\bibitem{42}C. Almeida, Am. J. Phys. 52 (1984) 921.
\bibitem{43}G. Kaiser, J. Math. Phys. 18 (1977) 952.
\bibitem{44}S. Twareque Ali and J.-P. Antoine, Ann. Inst. H. Poincar\'e 51 (1989) 23.
\bibitem{45}S. Twareque Ali, J.-P. Antoine and J.-P. Gazeau, Ann. Phys. 222 (1993) 38.
\bibitem{46}K. Kowalski, J. Rembieli\'nski and L.C. Papaloucas, J. Phys. A 29 (1996) 4149.
\bibitem{47}K. Kowalski and J. Rembieli\'nski, J. Phys. A 33 (2000) 6035.
\bibitem{48}K. Kowalski and J. Rembieli\'nski, Phys. Rev A 75 (2007) 052102.
\bibitem{49}K. Kowalski and J. Rembieli{\'n}ski, J. Phys. A: Math. Theor. 41 (2008) 304021.
\bibitem{50}F.W.J. Olver (Ed.), NIST Handbook of Mathematical Functions, 
Cambridge University Press, Cambridge, 2010.
\bibitem{51}I.S. Gradshteyn and I.M. Ryzhik, Tables of Integrals, Series, 
and Products, Elsevier, Amsterdam, 2007.
\bibitem{52}B. Rosenstein and M. Usher, Phys. Rev. D 36 (1987) 2381.
\bibitem{53}B. Rosenstein and L.P. Horwitz, J. Phys. A 18 (1985) 2115.
\bibitem{54}F.W. Strauch, Phys. Rev. A 73 (2006) 054302.
\bibitem{55}M.H. Al-Hashimi and U.-J. Wiese, Ann. Phys. 324 (2009) 2599.
\bibitem{56}T. Thiemann, Classical Quantum Gravity  13 (1996) 1383.
\bibitem{57}A.P. Prudnikov, Yu. A. Brychkov and O.I. Marichev, Integrals 
and Series. Vol 1. Elementary functions, Fizmatlig, Moscow, 2002.
\bibitem{58}A. Perelomov, Generalized Coherent States and Their Applications, Springer, Berlin, 1986.
\bibitem{59}K. Kowalski and J. Rembieli{\'n}ski, J. Math. Phys. 42 (2001) 4138.
\end{thebibliography}
\end{document}